\documentclass[ aip,
 apl,amsmath,amssymb,
 reprint,]{revtex4-1}
\usepackage{graphicx}
\usepackage{dcolumn}
\usepackage{bm}

\begin{document}
\title{Quantum simulation of general semi-classical Rabi model beyond strong
driving regime}
\author{Kunzhe Dai}
\thanks{These authors contributed equally to this work}
\affiliation{National Laboratory of Solid State Microstructures, \\
School of Physics, Nanjing University, Nanjing 210093, China}
\author{Haiteng Wu}
\thanks{These authors contributed equally to this work}
\affiliation{National Laboratory of Solid State Microstructures, \\
School of Physics, Nanjing University, Nanjing 210093, China}
\author{Peng Zhao}
\affiliation{National Laboratory of Solid State Microstructures, \\
School of Physics, Nanjing University, Nanjing 210093, China}
\author{Mengmeng Li}
\affiliation{National Laboratory of Solid State Microstructures, \\
School of Physics, Nanjing University, Nanjing 210093, China}
\author{Qiang Liu}
\affiliation{National Laboratory of Solid State Microstructures, \\
School of Physics, Nanjing University, Nanjing 210093, China}
\author{Guangming Xue}
\affiliation{Synergetic Innovation Center of Quantum Information $\&$ Quantum Physics, \\
University of Science and Technology of China, Hefei, Anhui 230026, China}
\author{Xinsheng Tan}
\email{txs.nju@gmail.com}
\affiliation{National Laboratory of Solid State Microstructures, \\
School of Physics, Nanjing University, Nanjing 210093, China}
\author{Haifeng Yu}
\email{hfyu@nju.edu.cn}
\affiliation{National Laboratory of Solid State Microstructures, \\
School of Physics, Nanjing University, Nanjing 210093, China}
\affiliation{Synergetic Innovation Center of Quantum Information $\&$ Quantum Physics, \\
University of Science and Technology of China, Hefei, Anhui 230026, China}
\author{Yang Yu}
\affiliation{National Laboratory of Solid State Microstructures, \\
School of Physics, Nanjing University, Nanjing 210093, China}
\affiliation{Synergetic Innovation Center of Quantum Information $\&$ Quantum Physics, \\
University of Science and Technology of China, Hefei, Anhui 230026, China}
\date{\today}

\begin{abstract}
We propose a scheme to simulate the interaction between a two-level system
and a classical light field. Under the transversal driving of two microwave
tones, the system Hamiltonian is identical to that of the general
semi-classical Rabi model. We experimentally realize this Hamiltonian with a
superconducting transmon qubit. By tuning the strength, phase and frequency
of the two microwave driving fields, we simulate the quantum dynamics from
weak to extremely strong driving regime.  The resulting evolutions gradually deviate from the
normal sinusoidal Rabi oscillations with increasing driving strength,
in accordance with the predictions of the general semi-classical Rabi model far beyond the weak driving limit. Our
scheme provides an effective approach to investigate the extremely strong
interaction between a two-level system and a classical light field. Such strong interactions are
usually inaccessible in experiments.
\end{abstract}

\maketitle

\preprint{AIP/123-QED}

Light-matter interaction has been at the heart of important modern
discoveries of fundamental effects, both classical and quantum mechanical
\cite{cohen1992atom}. As the simplest form of light-matter
interaction, a two-level atom interacting with a classical light field,
which was introduced by I. I. Rabi \cite{rabi1937space}, has been actively
investigated to study and control various quantum systems, including that of nuclear
magnetic resonance \cite{vandersypen2005nmr}, cavity quantum electrodynamics
(QED) \cite{haroche2006exploring}, and circuit QED \cite%
{blais2007quantum,you2003quantum}. Based on this semi-classical model \cite%
{braak2016semi}, Jaynes and Cummings introduced a fully quantized version,
i.e. quantum Rabi model \cite{jaynes1963comparison,tavis1968exact}, which
describes a two-level atom interacting with a quantum field mode. This model
is a cornerstone of various areas of quantum physics such as quantum optics
and quantum information processing.

As per the usual description of the Rabi model (semi-classical or quantum), the interaction between the atom and the field mode (classical or
quantum) can be decomposed into two parts, i.e. the rotating and the
counter-rotating terms. Usually, the coupling is significantly smaller than the atomic transition frequency and field mode frequency. In this weak-coupling regime, the dynamics of
the atom-field system is well described by the Rabi model with the usual
rotating wave approximation (RWA) \cite{jaynes1963comparison,tavis1968exact}%
, i.e. omitting the counter-rotating terms. However, when the
coupling is comparable to the atomic transition frequency and
field mode frequency, a limit which can be dubbed ultrastrong coupling (USC) regime, the
usual RWA breaks down \cite{bloch1940magnetic}, consequently the counter-rotating
terms manifest important impact on the dynamics of the atom-field
system\cite{x2011a,ashhab2010qubit,zueco2009qubit}. Xie \textit{et al.} \cite{xie2014anisotropic} have introduced an
anisotropic version of the quantum Rabi model thus generalized the quantum
Rabi model to USC regime. The experimental explorations of light-matter
interaction in ultra-strong coupling regime has long been restricted by the
coupling strength between the atom and field. Recently, a few experiments
have reached USC regime in solid state system \cite%
{fuchs2009gigahertz,laucht2016breaking,rao2017nonlinear,
forn2010observation,tuorila2010stark,deng2015observation,niemczyk2010circuit,fedorov2010strong, todorov2010ultrastrong,anappara2009signatures,yoshihara2017characteristic}
and have demonstrated complex dynamics unique to this regime, e.g. the Bloch-Siegert shift
\cite{forn2010observation,tuorila2010stark}, the Floquet state \cite%
{deng2015observation}, exotic behaviors distinct from the normal sinusoidal Rabi
oscillation \cite{fuchs2009gigahertz,laucht2016breaking,rao2017nonlinear},
and the spectroscopic signatures of these driving regimes \cite%
{niemczyk2010circuit,fedorov2010strong,todorov2010ultrastrong,anappara2009signatures,yoshihara2017characteristic}%
. However, it still remains an interesting but tough object to achieve driving strength that is significantly larger than the atomic transition frequency and field mode frequency (deep coupling regime \cite{casanova2010deep}%
). As an alternative to access this fascinating
field, quantum simulation of light-matter interaction in deep strong
coupling (for quantum Rabi model) or extremely strong driving (for semi-classical Rabi model) would be helpful for researching and/or
understanding various associated effects \cite%
{ballester2012quantum,grimsmo2013cavity,pedernales2015quantum,langford2016experimentally,fedortchenko2017quantum,puebla2017,braumuller2016analog}%
.

In this paper, first we propose a scheme to simulate the general semi-classical
Rabi model, for which the rotating and counter-rotating terms
have two different driving strengths. By using bichromatic driving, we
can engineer the desired effective Hamiltonian describing the generalized
semi-classical Rabi model in various driving regimes. Then we experimentally
realize the effective Hamiltonian in a circuit-QED setup, where a superconducting
transmon qubit is driven by two phase-locked microwave tones
simultaneously. We observe the dynamics of this light-matter interaction system
in different driving regimes. The population evolutions gradually deviate from the
normal sinusoidal Rabi oscillations with increasing driving strength.
From weak to extremely strong driving regimes, the system's behavior can be well
described by the general semi-classical Rabi model. Our scheme thus provides an
effective approach to explore the extremely strong interaction between a
two-level system and a classical light field, such strong interaction is usually inaccessible in experiments.

The Hamiltonian of the generalized semi-classical Rabi model is given as
\cite{xie2014anisotropic}
\begin{equation}
\begin{aligned} \hat
H/\hbar &=\frac{\omega_{a}}{2}{\hat\sigma}_{z}+A_{d}(H_{r}+\lambda H_{cr}),
\\{\hat
H}_{r}/\hbar &=e^{i\omega_{d}t}{\hat\sigma}_{-}+e^{-i\omega_{d}t}{\hat%
\sigma}_{+}, \\{\hat
H}_{cr}/\hbar &=e^{-i\omega_{d}t}{\hat\sigma}_{-}+e^{i\omega_{d}t}{\hat%
\sigma}_{+}. \end{aligned}
\end{equation}%
where ${\hat{\sigma}}_{z}$ is the Pauli matrices, ${\hat{\sigma}}_{-}({\hat{%
\sigma}}_{+})$ are the ladder operators for the atom of frequency $\omega
_{a}$, $A_{d}$ is the driving strength of the rotating interaction ${\hat{H}}%
_{r}$, and $\lambda $ is the relative strength between the rotating terms ${\hat{H}%
}_{cr}$ and counter-rotating terms ${\hat{H}}_{r}$. When $\lambda =1$, the
Hamiltonian reduces to that of the usual semi-classical Rabi model. When $\lambda
\neq 1$, this generalization can be considered as a semi-classical version
of the anisotropic Rabi model \cite{xie2014anisotropic},or as the anisotropic
generalization of the semi-classical Rabi model.

In order to reach the strong driving regime, we employ a bichromatic
driving, rather than a single large amplitude\cite{deng2015observation}, to drive the
two-level system. The Hamiltonian under the bichromatic driving in the laboratory frame with RWA can be written as
\begin{equation}
\begin{aligned} \hat H/\hbar = &\frac{{\omega _a}}{2}{\hat \sigma _z} +
\frac{{{\Omega_{1}}}}{2} \bigl ({e^{ + i({\omega _1}t + {\varphi _1})}}{\hat \sigma
_ - } + \text{H.c.}\bigr)\\ &+ \frac{{{\Omega_2}}}{2}\bigl ({e^{ + i({\omega _2}t + {\varphi
_2})}}{\hat \sigma _ - } + \text{H.c.} \bigr). \end{aligned}
\end{equation}%
with ${\Omega _{i}}$ the driving strength and ${\omega _{i}}$ the
frequency of the $i\left( {1,2}\right)$th drive. ${\varphi _{i}}$ denotes
the relative phase of the $i$th driving in the coordinate system of the qubit
Bloch sphere in the laboratory frame.

Then we transform from the laboratory frame to a reference frame that rotates
about the $z$ axis using the unitary transformation, $U_1= \exp[-i \omega t\sigma_z/2]$, with ${\omega }=({\omega _{1}}+{\omega _{2}})/2$. Finally we apply a time-independent rotation, $U_{2}= \exp[-i\phi\sigma _z/2]$, with $\phi =({\varphi _{1}}+{\varphi _{2}})/2$. Since $%
H_{\text{eff}}=U^{-1}HU-iU^{-1}\dot{U}$, after the above two steps the original Hamiltonian
in Eq. (2) becomes,
\begin{equation}
\begin{aligned} \hat H_{\text{eff}}/\hbar = &\frac{{\omega _a^*}}{2}{\hat \sigma
_z} + \frac{\Omega_{1}}{2}({e^{ + i(\omega _d^*t + \varphi _0^*)}}{\hat
\sigma _ - } + \text{H.c.})\\ &+ \frac{\Omega_{2}}{2}({e^{ - i(\omega _d^*t +
\varphi _0^*)}}{\hat \sigma _ - } + \text{H.c.}). \end{aligned}
\end{equation}%
Here we find that the effective qubit energy splitting $\omega _{a}^{\ast
}\equiv {\omega _{a}}-{\omega }$, the effective light field frequency $%
\omega _{d}^{\ast }\equiv \frac{1}{2}({\omega _{1}}-{\omega _{2}}),$ $\omega
_{1}>\omega _{2}$, and the initial phase $\varphi _{0}^{\ast }=({\varphi _{1}%
}-{\varphi _{2}})/2$. The driving regime is now completely determined by the
tunable parameters $\omega _{a}^{\ast },$ $\omega _{d}^{\ast },$ ${\Omega
_{1}}\mathrm{,}$ and ${\Omega _{2}.}$ Under the condition ${\Omega
_{1}=\Omega _{2}=\Omega ^{\ast }},$ the effective Hamiltonian can be reduced
to
\begin{equation}
\begin{aligned} \hat H_{\text{eff}}/\hbar &= \frac{{\omega _a^*}}{2}{\hat \sigma
_z}+{\Omega ^*}\cos(\omega _d^*t + \varphi _0^*){\sigma _x}. \end{aligned}
\end{equation}%
Eq. (4) is identical to the semi-classical Rabi Hamiltonian. It is interesting to
observe the close resemblance between the formalism of the effective Hamiltonian in
Eq. (3) and that of the Hamiltonian given in Eq. (1). This implies that we
can simulate the generalized semi-classical Rabi model with controllable
parameters. For a qubit with fixed ${\omega _{a},}$ we can tune $\omega $
and $\Omega _{1,2},$ thus changing the system from weak driving, to
ultrastrong driving $\Omega _{1,2}>0.1\omega _{a}^{\ast }$, then deep strong
driving $\Omega _{1,2}>\omega _{a}^{\ast }$, and even extremely strong
driving $\Omega _{1,2}\gg \omega _{a}^{\ast }.$ By choosing different
parameters, we can investigate the behaviors in anisotropic Rabi model,
phase sensitivity, and bias-modulated dynamics \cite%
{LU2016BIAS,oliver2005mach-zehnder,ashhab2007two} in various driving regimes.

\begin{figure}[tbph]
\includegraphics[width=8cm,height=5cm]{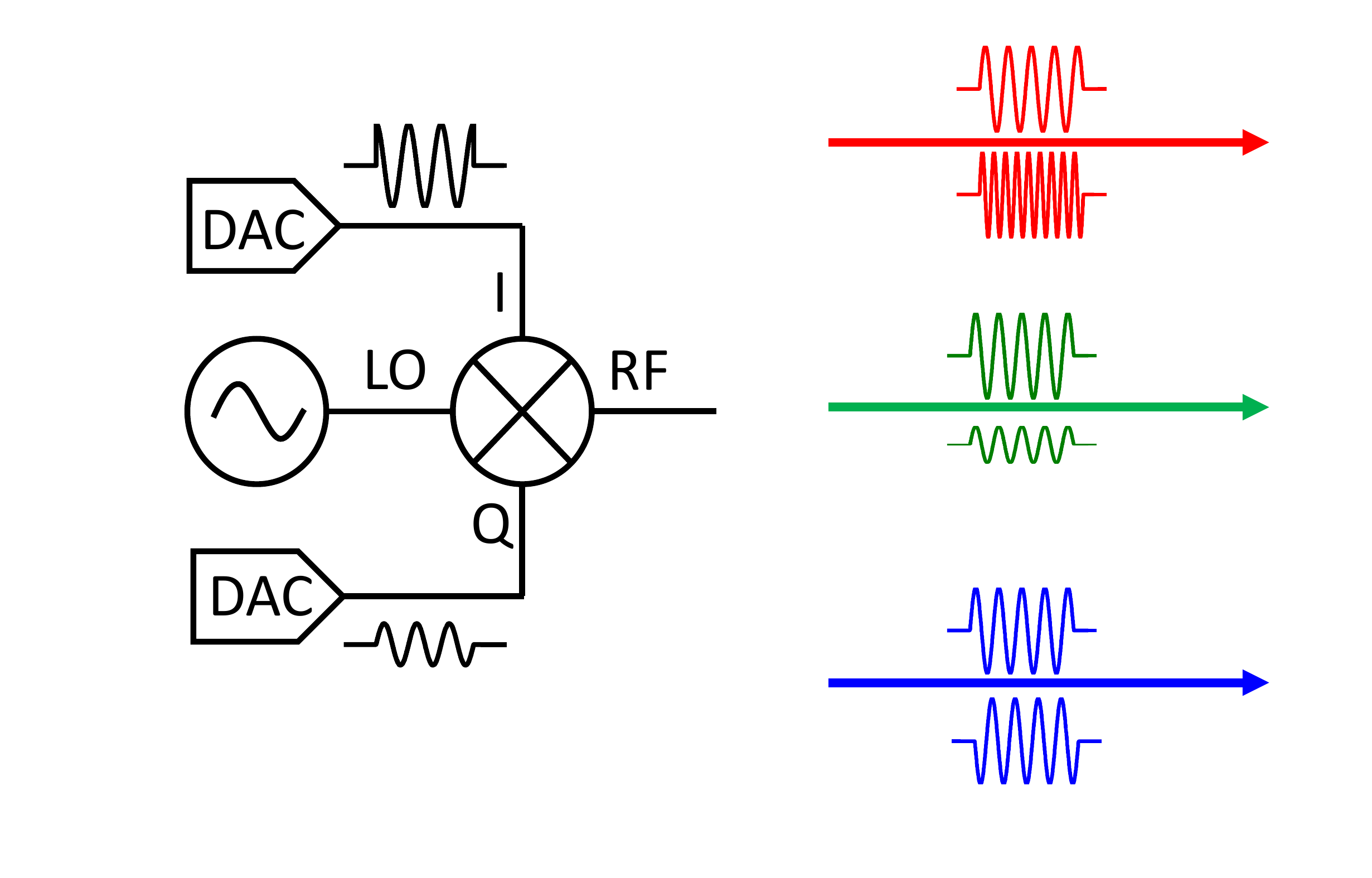}\newline
\caption{(Color Online) The schematic of the bichromatic driving scheme, capable of generating two distinct microwave tones, each with its own tunable frequency, strength, and phase, as illustrated by different colors in the figure.}
\end{figure}

\begin{figure*}[tbh]
\includegraphics[width=16cm,height=8.0cm]{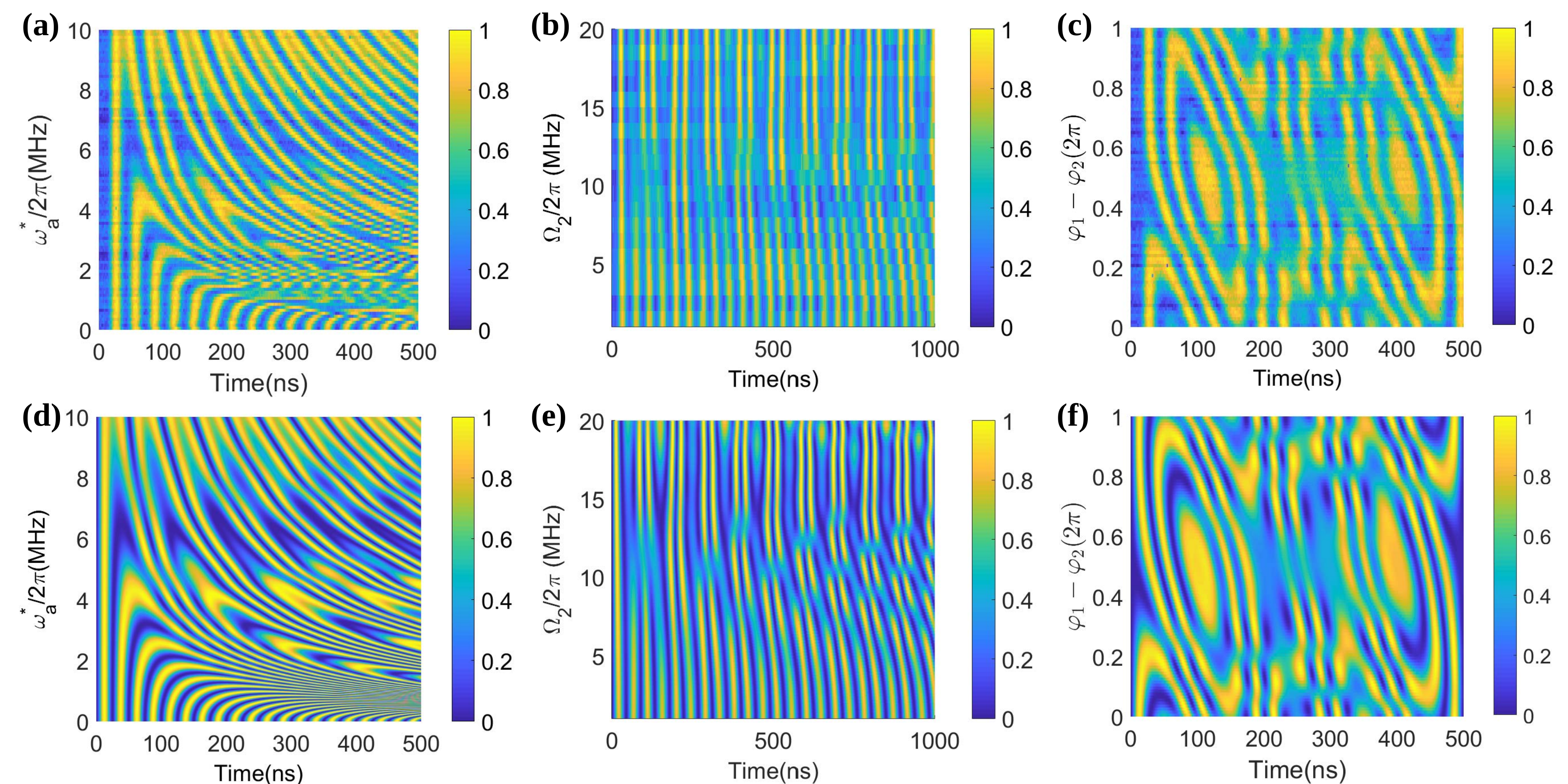}\newline
\caption{(Color Online) Qubit evolution in strong driving regime. (a), (b), and (c) are
measured qubit population as a function of time for tuning different
parameters. (d), (e), and (f) are the corresponding numerical simulations
with Hamiltonian in Eq. (3). }
\end{figure*}

We use a 3D transmon qubit to experimentally implement the quantum
simulation. A 3D transmon qubit comprises a superconducting transmon qubit and an aluminum holder in which the qubit sits, forming a typical circuit QED system. By our design, the detuning between the qubit frequency and cavity frequency is much
larger than the coupling strength between the qubit and the cavity, thus the
system works in the dispersive region. The main purpose of the cavity
in our experiments is to serve as a convenient tool to manipulate and
measure the qubit. The sample is cooled in a cryogen-free dilution
refrigerator to a base temperature of about $20~\mathrm{mK}$. The details of qubit
control and measurement can be found elsewhere\cite{reed2010high}. From spectroscopy
measurement, we obtain the qubit transition frequency at $\sim 7.173~\mathrm{GHz}$, the frequency difference between the ground
state and the first excited state of the qubit. The
resonant frequency of the 3D cavity is $\sim 9.052~ \mathrm{GHz}$. The coupling
strength between the qubit and the cavity is about $50~\mathrm{MHz}$. The energy
relaxation time of the qubit is about $10~\mathrm{\mu s}$ and the decoherence time
measured from Ramsey experiment is about $10~\mathrm{\mu s}$.

In order to ensure phase control of the driving microwaves with respect to the
qubit Bloch sphere coordinate system, we use a single microwave source
together with two digital to analog converter(DAC) channels of an Arbitrary Wave Generator(AWG) of the model Tektronix 5014c to generate two
microwave tones. Our bichromatic driving scheme (shown in Fig. 1) is
different from that typical used in strong driving experiments \cite%
{braumuller2016analog}. We use only one microwave source for Local Oscillator(LO) input, with
frequency equal to the mean frequency of the two tones and amplitude fixed.
The two DAC channels synchronously generate a cosine and a sine
waveform of different amplitudes for heterodyne IQ (In-phase and Quadrature) mixing. Except for the difference of a quadrature, these two waveforms have the same frequency and
initial phase. Using this scheme, we can obtain two separate microwave tones
with fully controllable frequency, strength and phase, which have simple
relations to the input waveforms and can be readily deduced. The power of
microwave required in this experiment is less than $20~\mathrm{dBm}$ (corresponding to driving strength at $2\pi \times 50~\mathrm{MHz}$ under our experimental setup), and the frequency applied is comparable to
qubit transition frequency. These requirements are met by most
microwave sources from common brands.

\begin{table}[tbh]
\centering
\begin{tabular}{c|cccccccc}
\toprule Setting & $\omega_1-\omega_a$ & $\omega_2-\omega_a$ & $\Omega_1$ & $%
\Omega_2$ & $\varphi_1-\varphi_2$ & $\omega_a^*$ & $\omega_d^*$ & $%
\varphi_0^*$ \\ \hline
a & 0 & 0$\sim$20 & 20 & 20 & 0 & 0$\sim$10 & 0$\sim$10 & 0 \\
b & 0 & 10 & 20 & 1$\sim$20 & 0 & 5 & 5 & 0 \\
c & 0 & 10 & 20 & 20 & 0$\sim$$2\pi$ & 5 & 5 & 0$\sim$$\pi$ \\ \hline
\end{tabular}%
\caption{The parameters used for quantum simulations in strong driving
regime. The frequency and the driving strength are in the unit of 2$\protect%
\pi~\mathrm{MHz}$, while the phase in radians. We used the same parameters for numerical simulations.}
\end{table}

According to Eq. (3), we can individually tune multiple parameters to simulate quantum
dynamics in strong driving regime. In our experiments, we choose three
representative settings to fulfill the requirement of strong driving. Table I lists
the parameters used for different settings. At first, we keep the driving
strength constant, i.e. $\Omega _{1}/2 \pi=\Omega _{2}/ 2 \pi= 20~\mathrm{MHz}$, the model reduces
to Rabi model as described in Eq. (4). The initial phases of both microwave tones are the same, rendering $\varphi ^{\ast }=0$. As we keep the first microwave tone in resonance with the qubit and change the frequency of the second tone, we obtain different values for the
effective transition frequency $\omega _{a}^{\ast }$, hence also for the ratio $\Omega ^{\ast }/ \omega _{a}^{\ast }$, which scales the relative driving strength.
Shown in Fig. 2(a) is the qubit population at the first excited state as a function of time for
different driving strengths with $\omega _{a}^{\ast }/ 2 \pi$ adjusted from $10~\mathrm{MHz}$ to $0$, corresponding to the ratio $\Omega ^{\ast }/\omega
_{a}^{\ast }$ increasing from $2$ to infinity. The system in the process reaches and even
goes beyond the deep strong driving regime. The resulting oscillations are obviously
anharmonic and nonlinear, exhibiting the dynamics of semi-classical Rabi
model. In the second setting, we only change the driving strength of the
second microwave tone $\Omega _{2}$ and keep the other parameters fixed.
From Eq. (3) we can see that $\Omega _{2}$ defines the strength of the
counter rotating field. This setting can be used to investigate the
anisotropic Rabi model. We measure the qubit population evolution under various
values for $\Omega _{2}$, as shown in Fig. 2(b). When $\Omega _{2}$ is small,
corresponding to the bottom part of Fig. 2(b), distinctly sinusoidal oscillations
are observed because the the effects of the counter-rotating terms are negligible. With the
increase of $\Omega_2$, the sinusoidal fringes deform and transition into an irregular pattern, exhibiting the influence of the counter-rotating terms under
large $\Omega _{2}$. In the third setting, we only change the phase
difference of the two tones, which defines the initiate phase $\varphi ^{\ast }$
in the effective Hamiltonian in Eq. (3). In the weak driving regime this
initial phase has negligible effects on the dynamics of the system\cite{london2014strong,rao2017nonlinear}. However, this is not the case when we
set the parameters for strong driving regime and measure the corresponding qubit evolution, as shown in Fig. 2(c), in which the phase-sensitivity of the evolution dynamics is evident. We note that
the period of phase $\varphi^*_0$ for this pattern is $\pi $ instead of $2\pi $. In addition,
we provide numerical simulation results by solving the master equations. The
computed qubit evolutions agree with the experimental results.

\begin{figure}[tbp]
\includegraphics[width=7.0cm]{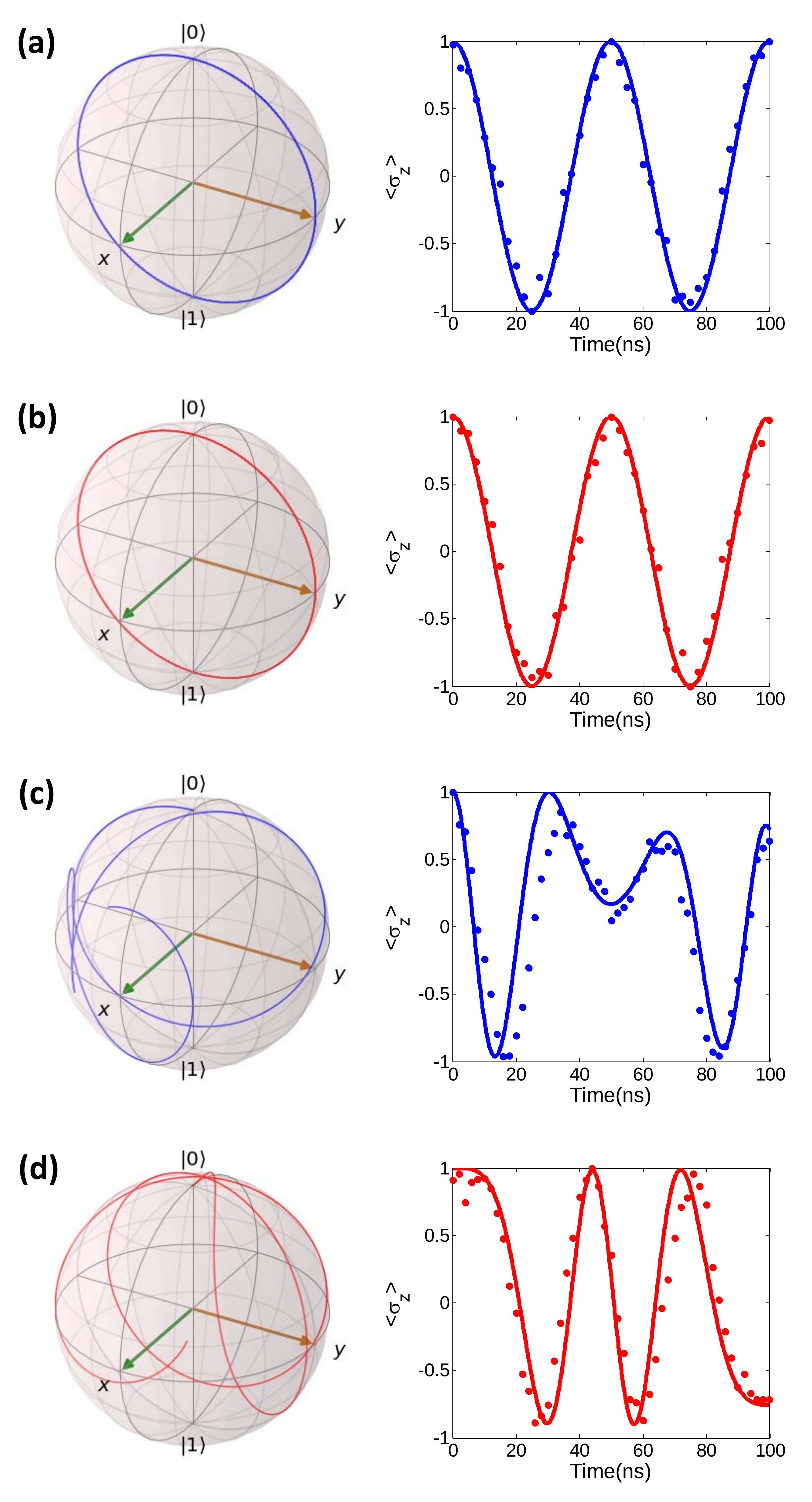}\newline
\caption{(Color Online) The time evolution dynamics in Bloch sphere representation (left)
and the corresponding $\langle \protect\sigma _{z}\rangle $ component
(right) of the state vector with different initial phases. (a) and (b) [(c) and (d)] are evolutions in weak (deep strong) driving limit, with initial phase
$\varphi_0^* = 0$ and $\pi /2$, respectively. Solid lines and symbols correspond to theoretical calculation and experimental data.}
\end{figure}
\ \

In strong driving regime, a remarkable feature is the presence of counter-rotating evolution, which interestingly shows the initial phase dependence
\cite{laucht2016breaking,london2014strong}. One of the advantages of our scheme
is precise control of initial phase. As a result, we can investigate the
dynamics with a specific initial phase instead of using an averaged phase
\cite{laucht2016breaking,london2014strong}. Fig. 3 shows the state evolution
on Bloch sphere (left) and its corresponding $\langle \sigma _{z}\rangle $
component (right) as a function of time. For simplicity, we choose on-resonance condition $\omega _{d}^{\ast }=\omega _{a}^{\ast }.$ By changing
the values of $\omega _{a}^{\ast }$ while keeping $\Omega ^{\ast }/ 2 \pi =20~\mathrm{MHz}$, we simulate the dynamics from weak driving to deep strong driving regime.
For  (a) [(c)] and (b) [(d)], $\varphi _{0}^{\ast }=0$
and $\pi /2,$ respectively. In Fig. 3(a), (b), $\omega _{a}^{\ast }/2 \pi= 2~\mathrm{GHz} $,
which is much larger than $\Omega ^{\ast }/2 \pi$. The system is in weak driving
regime, where RWA is valid. The qubit state vector rotates normally in
the $y$\nobreakdash--$z$~plane on the surface of the Bloch sphere and the trajectories of both
$\varphi _{0}^{\ast }$'s coincide. In Fig. 3(c), (d), $\omega _{a}^{\ast }/2 \pi=5~\mathrm{MHz} $, and $\Omega ^{\ast }>\omega _{a}^{\ast }$, corresponding to the deep
strong driving regime. Two trajectories completely deviate from the $y$\nobreakdash--$z$~plane,
the $\langle \sigma _{z}\rangle $ component of state vector shows
complicated oscillations. The trajectories of state vector for two initial
phases are completely different. The dynamics exhibits strong sensitivity to
initial phase. These phenomena result from the significant contribution of
counter-rotating terms. In addition, the oscillations of $\langle \sigma
_{z}\rangle $ component agree with the numerical simulations, indicating
precise control of the initial phase realized by our scheme.

In summary, we propose a bichromatic driving method to simulate
the general Rabi model in strong driving regime. This method allows us to
reach the strong driving regime with easily attainable experimental
conditions. We demonstrate the dynamics of general Rabi model in various driving regimes. When the driving strength is higher than the
effective energy splitting $\omega_a^*$, the model exhibits highly anharmonic and
nonlinear behaviors, clearly deviating from the usual sinusoidal Rabi
oscillations. Such effects manifest the influence of the counter-rotating terms upon the breakdown of RWA. Moreover, these sophisticated
patterns and trajectories, obtained experimentally, are in agreement with the computed results. This quantum simulation scheme provides a useful test bed for studying
general Rabi model in regimes defined by arbitrarily strong driving, and for exploring optimal quantum
gate operations.

This work was partly supported by the the NKRDP of China (Grant No. 2016YFA0301802), NSFC (Grant No. 11504165, No. 11474152, No. 61521001).

\end{document}